\documentclass[12pt]{article}
\usepackage[dvips]{graphicx}
\usepackage{epsfig}
\usepackage{latexsym}

\newcommand{\ice}[1]{\relax}
 
\newcommand{\beq}{\begin{equation}}
\newcommand{\eeq}{\end{equation}}
\newcommand{\be}{\begin{equation}}
\newcommand{\ee}{\end{equation}}
\newcommand{\ba}{\begin{eqnarray}} 
\newcommand{\ea}{\end{eqnarray}} 
 
\newcommand{\GeV}{{\rm\,GeV}}
\newcommand{\MeV}{{\rm\,MeV}}

\newcommand{\MSbar}{{\overline{\rm MS}}}
\newcommand{\pfrac}[2]{\left(\frac{#1}{#2}\right)}
\newcommand{\slq}{q\kern-5.5pt/}
\newcommand{\slu}{u\kern-5.5pt/}
\newcommand{\slx}{x\kern-5pt\raise1pt\hbox{$\scriptstyle/$}\kern1pt}
\newcommand{\eps}{\varepsilon}

\begin{document}
\thispagestyle{empty}
\begin{flushright}
MZ-TH/07-20\\
arXiv: 0805.3590
\end{flushright}
\vspace{0.5cm}

\begin{center}
{\Large \bf Next-to-Leading Order perturbative QCD 
corrections to baryon correlators in matter}\footnote[4]{Supported in part by
the RFFI grant No.~06-02-16353a.}\\[0.5truecm]
{\large S.~Groote,$^{1,2}$ J.G.~K\"orner$^1$ and
  A.A.~Pivovarov$^{1,3}$}\\[.4cm]
$^1$ Institut f\"ur Physik, Johannes-Gutenberg-Universit\"at,\\[.1truecm]
  Staudinger Weg 7, D-55099 Mainz, Germany\\[.3truecm]
$^2$ Teoreetilise F\"u\"usika Instituut, Tartu \"Ulikool,
  T\"ahe 4, 51010 Tartu, Estonia\\[.3truecm] 
$^3$ Institute for Nuclear Research of the\\[.1truecm]
  Russian Academy of Sciences, Moscow 117312, Russia
\end{center}

\begin{abstract}
We compute the next-to-leading order (NLO) perturbative QCD corrections to the
correlators of nucleon interpolating currents in relativistic nuclear matter.
The main new result is the calculation of the ${\cal O}(\alpha_s)$ 
perturbative corrections to the coefficient functions of the vector quark 
condensate in matter. This condensate appears in matter due to the violation of
Lorentz invariance. The NLO perturbative QCD corrections turn out to be large
which implies that the NLO corrections must be included in a sum rule 
analysis of the properties of both bound nucleons and relativistic 
nuclear matter.
\end{abstract}

\newpage

\section{Introduction}
The study of bound states in QCD is a difficult problem. With more than 30
years of research it is clear that the most promising approach to obtain
quantitative information on the properties of hadrons are very likely
lattice techniques in particular since both computer power and computational 
methods advanced dramatically since their first introduction in the early 
seventies. Lattice results are now available in many hadronic channels and
further research is being actively pursued~\cite{Lepage:2004mq}. Nevertheless
analytical nonlattice techniques can be used to verify at least the
consistency of some models for hadron description and their predictions. The
QCD sum rule analysis is based on the operator product expansion (OPE) and
serves as a rigorous framework for many calculations in the theory of
hadrons~\cite{Shifman,Reinders,Krasnikov,AlikPietro-SumRules}. QCD sum rules
are also useful for testing some model dependent
approaches~\cite{Adkins:1983ya,Diakonov:1997mm,Braaten:1989rg} such as the MIT
quark-bag model~\cite{Chodos}. A further important problem is the quantitative
description of the properties of bound nucleons and relativistic nuclear
matter within QCD.

The proton is the most abundant strongly interacting particle on Earth. It has
played an important part in particle physics since long ago. In the theory of
strong interactions the proton is a bound state of quarks and gluons with
three valence quarks fixing its discrete quantum numbers. It has been
intensively studied during the last fifty years. At present one cannot
directly compute the properties of the proton analytically from QCD even for
an isolated proton. The techniques of the QCD sum rules provides a powerful
tool for the phenomenological analysis of the proton. The properties of
baryons have been successfully described within this approach when account is
taken of the leading vacuum condensates~\cite{Ioffe:1981kw,Chung,Kras1}. The
ever improving accuracy of experimental data requires improvements also in the
theoretical description. Primarily this means that one has to account for the
perturbative QCD corrections to the coefficient functions of the OPE for
baryonic correlators. The next-to leading order (NLO) perturbative QCD
corrections have been calculated for the coefficient functions of the OPE for
the unit operator and the scalar quark condensate in the massless quark limit
in~\cite{ba4}. The perturbative corrections have been found to be large. The
results for the unit operator were generalized to the massive quark case
in~\cite{barH} where again large perturbative corrections were found. The
properties of the proton in vacuum are well studied in QCD although there is
room for improvements in the numerical accuracy.

Protons are traditionally used as targets in accelerator experiments as e.g.\
in the electron scattering on iron at DESY. To analyze the data obtained in
these experiments one needs to know the properties of the protons bound in
nuclei, or more generally of the nuclear medium. Thus there is a considerable
interest in computing the parameters of the proton medium. The most obvious
reason is that protons are part of the nuclei which serve as targets in
accelerator experiments. The scattering on nuclei is different from the
scattering on the proton, and this is important for the interpretation of the
data. One of the best known examples is the EMC
effect~(e.g.~\cite{Akulinichev:1985ij}). One of the possible theoretical
approaches is to use an effective theory where the proton in the medium is
treated as an effective particle~\cite{Kulagin:kb,Meissner:2004yy,%
Bijnens:2005mi}. Another approach is to analyze the properties of nucleon
medium within the QCD sum rule approach in~\cite{drukarev}. In this approach
the problem of an accurate determination of the parton distributions in nuclei
ultimately requires the calculation of perturbative corrections to the OPE in
the framework of QCD sum rules. 

In the present paper we compute NLO perturbative QCD corrections to the
correlators of baryon interpolating currents in matter. We focus on the
calculation of the ${\cal O}(\alpha_s)$ perturbative corrections to the
coefficient functions of the bilinear quark operators that may lead to the
emergence of nonvanishing condensates upon averaging over the appropriate
physical states. We present new results for the coefficient functions of the
quark operators in the vector representation $(1/2,1/2)$ of the Poincare
group: NLO accuracy is achieved in the expansion in the coupling constant of
QCD. When averaged over the ground state of matter one obtains nonvanishing
values of the quark operators in the vector representation $(1/2,1/2)$ when
studying the properties of the nucleon medium within the QCD sum rules
approach. The presence of matter violates Lorentz invariance and thus allows
for the appearance of a vector condensate averaged over the matter states. The
NLO perturbative QCD corrections to the coefficient functions turn out to be
large in the $\MSbar$ scheme. This means that one must account for the
perturbative corrections in applications of sum rules analysis' of baryon
properties in matter.


\section{Basic expressions for the analysis}


The formulation of the OPE analysis is standard by now. In accordance with the 
QCD sum rule approach, we shall calculate the operator product expansion of 
two interpolating currents $J(x)$ which have a nonvanishing overlap with the
state of interest. The OPE for the quantity
\begin{equation}
T(q)=i\int d^4xe^{iqx}T\{J(x)\bar J(0)\}
\end{equation}
is performed by means of Wilson's operator product expansion. For the analysis
of the properties of isolated hadrons in the vacuum one then averages the
operator product over the ground state of QCD or the physical vacuum to obtain
the correlation function of two interpolating currents $J(x)$
\begin{equation}
\Pi(q)=i\int d^4xe^{iqx}\langle 0|T\{J(x)\bar J(0)\}|0\rangle 
\end{equation}
in vacuum. The assumption of the QCD sum approach is that the vacuum 
expectation values of the local operators that appear in the OPE (the so-called
condensates) are nonzero. The calculations are done in perturbative QCD and
therefore must be performed in the region $-q^2\ge 1\GeV^2$ where perturbative
QCD is valid. However, in this region, the effective strong interaction
constant $\alpha_s$ is not very small numerically~\cite{Korner:2000xk}. This
forces one to calculate the coefficient functions of the operator product
expansion in perturbation theory at least up to NLO in order to have sensible
results. One more reason is of course the general property of perturbation
theory that only at this order of the perturbative expansion can one reliably
fix the renormalization group scale $\mu$ which determines the numerical
values of the coupling constant and condensates within the OPE. It turns out
that the NLO corrections  are rather large in many hadronic channels. Even in
the case of the correlators of the classical quark--antiquark currents the
perturbative QCD corrections in the standard $\MSbar$ scheme are not
small~\cite{Baikov:2004tk}. Thus, the calculations of $\Pi(q)$ should be done
at least at NLO in $\alpha_s$ to have the precision required by modern
applications. 

In the original paper of Ioffe the current
\begin{equation}\label{ioffe1}
\eta=\epsilon_{abc}(u_a^TC\gamma_\mu u_b)\gamma_5\gamma^\mu d_c
\end{equation}
was used to analyze the properties of the proton~\cite{Ioffe:1981kw}. $u$ and 
$d$ are light quark fields and $C$ is the charge conjugation matrix with the
properties $C\gamma_\mu^TC=\gamma_\mu$ and $C=-C^{-1}=-C^T=-C^\dagger$. By
using a Fierz transformation the Ioffe current $\eta(x)$ can also be rewritten
as a linear combination of the two current operators $O_1$ and $O_2$,
\begin{equation}
\label{ioffe2}
\eta(x)=2(O_1-O_2)
\end{equation}
where 
\begin{equation}
O_1=\epsilon_{abc}(u_a^TCd_b)\gamma_5u_c,\qquad
O_2=\epsilon_{abc}(u_a^TC\gamma_5d_b)u_c.
\label{o1o2}
\end{equation}
In fact, the operators $O_1$ and $O_2$ form a complete basis for the lowest
dimension interpolating currents of the proton with no derivative couplings.
We generalize Ioffe's current by writing a linear combination of the operators
$O_1$, $O_2$ of the form 
\begin{equation}\label{general}
J(x)=O_1+tO_2
\end{equation}
where $t$ is a mixing parameter.

\begin{figure}\begin{center}
\epsfig{figure=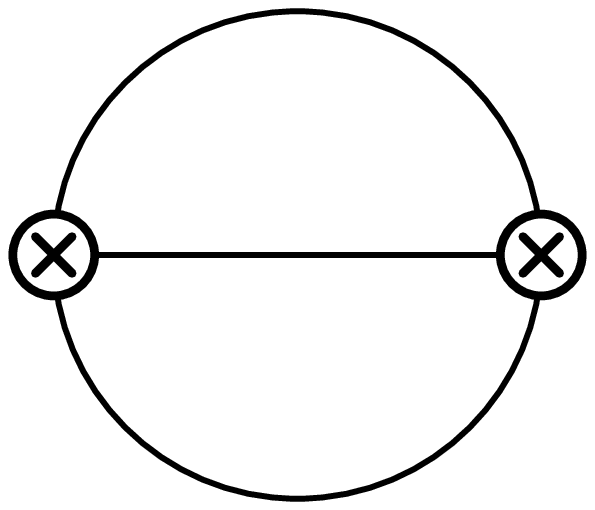, scale=0.5}\qquad
\epsfig{figure=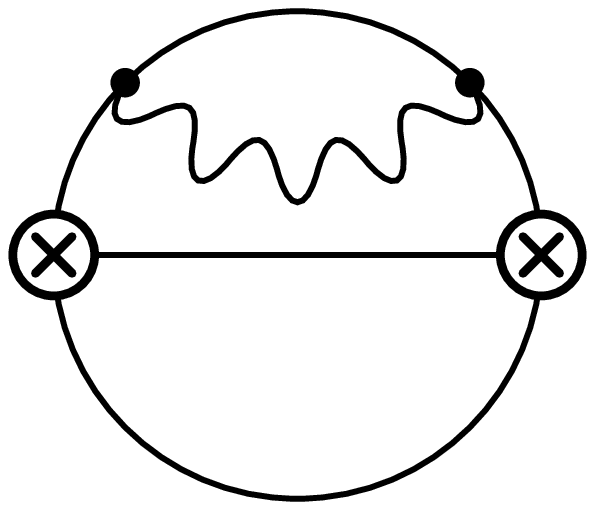, scale=0.5}\qquad
\epsfig{figure=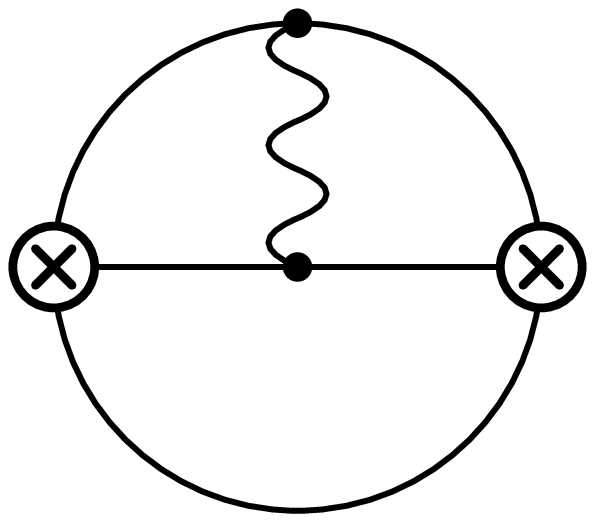, scale=0.5}\\[12pt]
(a)\kern92pt(b)\kern92pt(c)
\end{center}
\caption{\label{cond0}LO (a) and NLO propagator-type (b) as well as fish-type
(c) corrections of the baryonic two-point correlator.}
\end{figure}

The topology needed in the calculation of the LO correlator (see
Fig.~\ref{cond0}(a)) falls into the category of the well-known sunset diagrams.
These only contain lines that connect two vertices~\cite{Berends:1997vk,%
Groote:1998ic}. Such a topology also appears in the effective gluon low energy
correlator for heavy quarks below the production threshold that leads to the
decay of heavy quarkonia into gluon/photons~\cite{Groote:2001vr,%
Davydychev:1999ic}. Sunset diagrams can be calculated very efficiently in
configuration space. The NLO perturbative QCD corrections are of two types.
The first correction is a propagator-type correction which is not difficult to
compute (cf.\ Fig.~\ref{cond0}(b)). The second correction (Fig.~\ref{cond0}(c))
comes from the diagrams of the fish type and involves the calculation of an
irreducible two-loop subdiagram.

For completeness, and for the convenience of the reader, we present general
expressions for the corrections in configuration space which can be used for a
variety of interpolating currents. The efficiency of the configuration space
approach has already been proven in computing NLO corrections to pentaquark
correlators~\cite{Groote:2006sy}.

\begin{figure}\begin{center}
\epsfig{figure=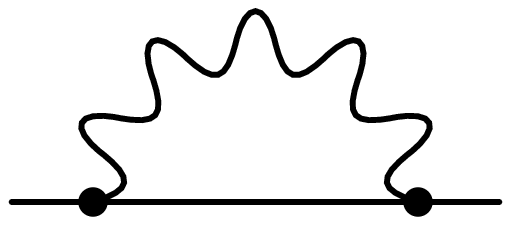, scale=0.5}\qquad
\epsfig{figure=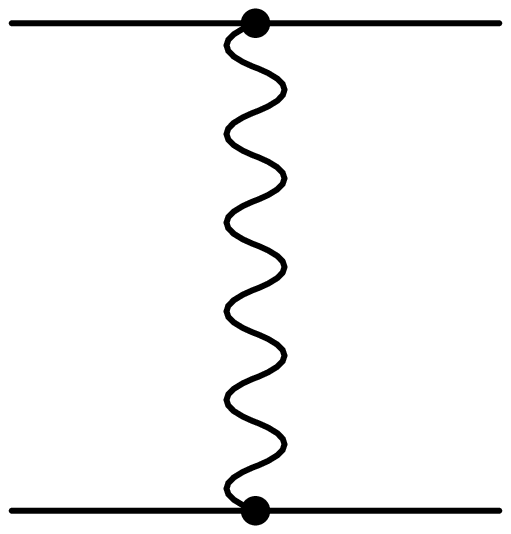, scale=0.5}\\[12pt]
(a)\kern80pt(b)\end{center}
\caption{\label{prodipro}propagator (a) and dipropagator (b) correction with
open Dirac indices}
\end{figure}

First we list the NLO correction to the fermion propagator (see
Fig.~\ref{prodipro}(a)) in configuration space. One has
\begin{equation}\label{propagator}
S(x)|_{\rm NLO}=S(x)|_{\rm LO}
\left\{1-C_F\frac{\alpha_s}{4\pi}
  \frac1\eps\left(\mu^2_Xx^2\right)^\eps\right\},
\end{equation}
where $C_F=(N_c^2-1)/2N_c$ is the Casimir operator of the color group $SU(N_c)$
($N_c=3$ for QCD) and $S(x)|_{\rm LO}=\gamma^\mu x_\mu F_0(x^2)$ is the LO
fermion propagator where $F_0(x^2)$ is defined in the Euclidean domain and is
given by
\begin{equation}
F_0(x^2)=\frac{-i\Gamma(2-\eps)}{2\pi^{2-\eps}(x^2)^{2-\eps}}.
\end{equation}
$\Gamma(z)$ is Euler's gamma-function. The space-time dimension is
parametrized by $D=4-2\eps$ throughout. Written in terms of $F_0(x^2)$ one has
\begin{equation}
S(x)|_{\rm NLO} =F_0(x^2)\gamma^\mu x_\mu
  \left\{1-C_F\frac{\alpha_s}{4\pi}\frac1\eps
  \left(\mu^2_Xx^2\right)^\eps\right\}.
\end{equation}
The renormalization scale $\mu_X$ is the appropriate one for calculations in
configuration space. This choice avoids the appearance of $\ln(4\pi)$ and
$\gamma_E$ (Euler constant) factors in configuration space calculations. The
relation of $\mu_X$ and the usual renormalization scale $\mu$ of the
$\MSbar$-scheme is given by $\mu_X=\mu e^{\gamma_E}/2$. Note that the NLO
fermion propagator is gauge dependent even if the complete calculation is
gauge invariant. In our calculation we have used diagonal or Feynman gauge.  

The next quantity needed in our calculation is the NLO correction to the
propagator of a pair of fermions. We call this diquark propagator a
dipropagator for short and denote it by $S_2(x)$. The dipropagator is given 
in terms of a two-loop amplitude with open Dirac indices (see
Fig.~\ref{prodipro}(b)) which requires a genuine two-loop calculation. The
result for the dipropagator up to NLO reads~\cite{Groote:2006sy} 
\begin{eqnarray}\label{dipropagator}
\lefteqn{S_2(x)|_{\rm NLO}
  \ =\ F_0(x^2)^2\{\gamma^\mu x_\mu \otimes\gamma^\nu x_\nu}\nonumber\\&&\strut
  +t^a\otimes t^a\frac{\alpha_s}{4\pi}\frac1\eps(\mu_X^2 x^2)^\eps
  (\gamma^\mu\otimes\gamma^\nu(a_1x_\mu x_\nu+b_1x^2g_{\mu\nu})
  \nonumber\\&&\strut\qquad\qquad\qquad\qquad
  +a_3\Gamma_3^{\alpha\beta\mu}\otimes{\Gamma_{3\ \alpha\beta}}^\nu
  x_\mu x_\nu)\}
\end{eqnarray}
where the coefficients $a_1$, $b_1$ and $a_3$ are given by 
\[
a_1=-1-\frac{11}2\eps,\quad
b_1=-1-\frac12\eps,\quad
a_3=-\frac12-\frac14\eps,
\]
and where
\begin{equation}
\Gamma_3^{\mu\alpha\nu}=\frac12(\gamma^\mu\gamma^\alpha\gamma^\nu
  -\gamma^\nu\gamma^\alpha\gamma^\mu).
\end{equation}
We use the standard notation $\otimes$ for the direct product of two Dirac
or color matrices. The generators of the color group algebra $t^a$ appearing
in the above expression are normalised by the condition
${\rm tr}(t^at^b)=\delta^{ab}/2$. Eqs.~(\ref{propagator})
and~(\ref{dipropagator}) allow one to calculate the NLO corrections to
$n$-quark(antiquark) current correlators of any composition using purely
algebraic algorithms without having to compute any integrals. For example, the
form~(\ref{dipropagator}) has been used in~\cite{Groote:2006sy} to compute the
radiative corrections to the pentaquark current correlator.  

The above results need to be renormalized. The renormalization can be done in
configuration space. To renormalize the single propagator one can use
multiplicative renormalization. The only ingredient needed is the wave function
renormalization constant of the fermion. The diagrams involving dipropagators
result in mixing of the operators under renormalization. Mixing is taken into
account through a subtraction of the corresponding vertex divergences
generated by the operator that can admix to the initial current. The general
formula reads
\begin{eqnarray}
\psi_i\otimes\psi_j|_R^{\rm IR}&=&\psi_i\otimes\psi_j
  -\frac{\alpha_s}{4\pi\eps}\left(1_{ii'}\otimes 1_{jj'}
  +\frac14\sigma^{\alpha\beta}_{ii'}\otimes\sigma^{\alpha\beta}_{jj'}\right)
  \psi_{i'}\psi_{j'}\nonumber\\
  &=&\psi_i\otimes\psi_j-\frac{\alpha_s}{4\pi\eps}\left(\psi_i\otimes\psi_j
  +\frac14\sigma^{\alpha\beta}_{ii'}\psi_{i'}\otimes\sigma^{\alpha\beta}_{jj'}
  \psi_{j'}\right),
\end{eqnarray}
where $i$ and $j$ are color indices and $\psi$ stands for either the up or
down quark fields. For definiteness we define our
$\sigma^{\alpha\beta}=i/2[\gamma^{\alpha},\gamma^{\beta}]$. The results are
again given in diagonal or Feynman gauge where we emphasize again that the
complete result is gauge independent. Note that the part proportional to the
product of $\sigma$-matrices is gauge independent.

Before presenting the results of our calculation we want to remark on the
renormalization group properties of the operators $O_{1,2}$ defined in
Eq.~(\ref{o1o2}). They represent a complete basis of the operators mixing
under renormalization and suffice to perform the calculation of the baryonic
correlators. As mentioned before the operators $O_{1,2}$ form a basis of
operators of lowest dimension for the interpolating currents of the nucleon.
Since their anomalous dimensions are identical at this order they satisfy the
same renormalization group evolution. One has 
\[
\mu^2\frac{d}{d\mu^2}O_{1,2}(\mu)=\frac{\alpha_s}{2\pi}O_{1,2}(\mu).
\]
Note that the numerical value of the anomalous dimension is such that the
product $\sqrt{m(\mu)}O_{1,2}(\mu)$ with $m(\mu)$ is renormalization group
invariant at this order of QCD. At NLO the operators mix. The two-loop
anomalous dimensions have been computed in Ref.~\cite{ba3}. These ingredients
allow one to compute the necessary correlator functions.


\section{Correlator including the scalar condensate}


In the OPE one computes the contributions of local operators to the correlator
function. In case of the vacuum correlators only Lorentz scalars contribute
which means that the only nonvanishing condensate is of the form
$\langle\bar qq\rangle$. The standard vacuum condensate contributions have
been calculated before in Refs.~\cite{ba4} including the
${\cal O}(\alpha_{s})$ corrections. They read
\begin{equation}
i\int dxe^{iqx}\langle T\{J(x)\bar J(0)\}\rangle=\slq\Pi_q(q^2)+\Pi_m(q^2)
\end{equation}
with
\[
\Pi_q(q^2) = -\frac1{8(4\pi)^4}(5t^2+2t+5)Q^4\ln\pfrac{Q^2}{\mu^2}
\left\{1+\frac{\alpha_s}\pi\left(\frac{71}{12}-\frac12\ln\pfrac{Q^2}{\mu^2}
\right)\right\}
\]
and
\begin{equation}\label{Pimq2}
\Pi_m(q^2)=\frac{\langle\bar\psi\psi\rangle}{4(4\pi)^2}Q^2
  \ln\pfrac{Q^2}{\mu^2}
  \Big(1-t\Big)\left(5+7t+(3+5t)\frac{3\alpha_s}{2\pi}\right).
\end{equation}
Here $Q^2=-q^2$, $\langle\bar\psi\psi\rangle=
\langle\bar uu\rangle=\langle\bar dd\rangle$.

In matter a new type of condensate
$\langle M|\bar q\gamma_\mu q|M\rangle\ne 0$ appears where $|M\rangle$ is the
ground state of the matter. The quantity
$\langle M|\bar q\gamma_\mu q|M\rangle$ violates Lorentz invariance. This is
expected since the matter itself fixes a special frame. Therefore one needs to
account for a new operator in the OPE for the baryonic currents. Including the
new vector operator one now has
\begin{equation}
T\{J(x)J(0)\}=C_{\rm I}(x^2)+C_{\bar qq}(x^2)\{\bar qq\}
  +C_{\bar q\gamma_\mu q}^\mu(x)\{\bar q\gamma_\mu q\}
\end{equation}
where the coefficient function $C_{\bar q\gamma_\mu q}^\mu(x)$ of the vector
operator $\{\bar q\gamma_\mu q\}$ is a four-vector. We calculate the
coefficient function $C_{\bar q\gamma_\mu  q}^\mu(x)$ at NLO accuracy by using
again configuration space techniques which were developed in a different
setting, namely the NLO analysis of pentaquark sum rules~\cite{Groote:2006sy}. 

\begin{figure}\begin{center}
\epsfig{figure=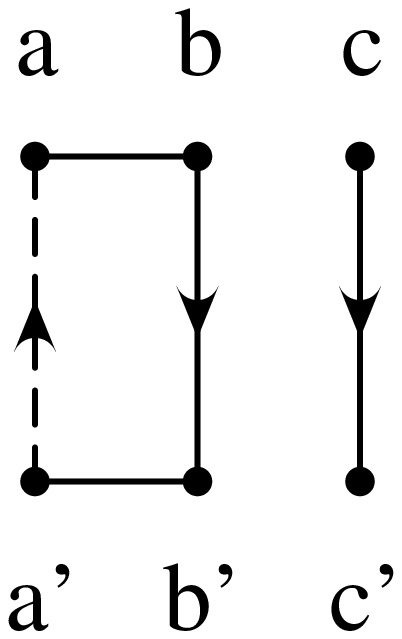, scale=0.5}\qquad\qquad
\epsfig{figure=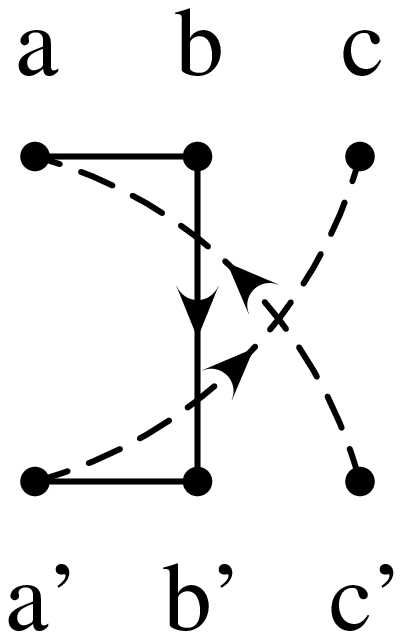, scale=0.5}\\[12pt]
(a)\kern80pt(b)\end{center}
\caption{\label{crossover}ordinary (a) and crossover part (b) of the Wick
contraction for the operators $O_{1,2}$ in a symbolic representation. The
arrows indicate the direction for the quark current within the correlator, the
dashed lines are inverted by using transposition and charge conjugation.}
\end{figure}

\begin{figure}\begin{center}
\epsfig{figure=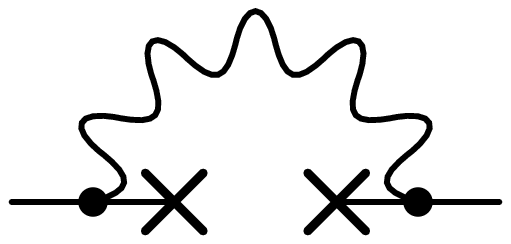, scale=0.5}\qquad\qquad
\epsfig{figure=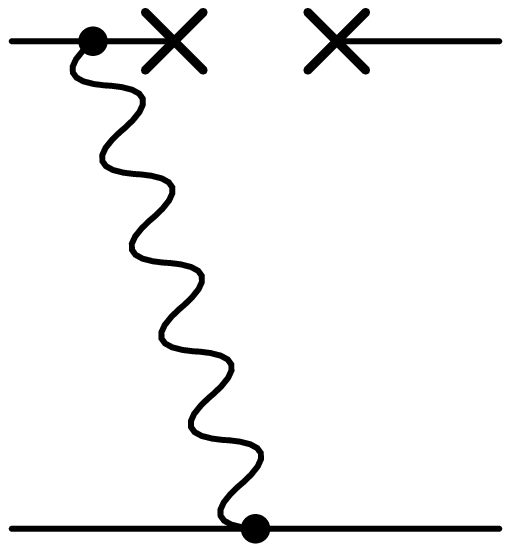, scale=0.5}\quad+\quad
\epsfig{figure=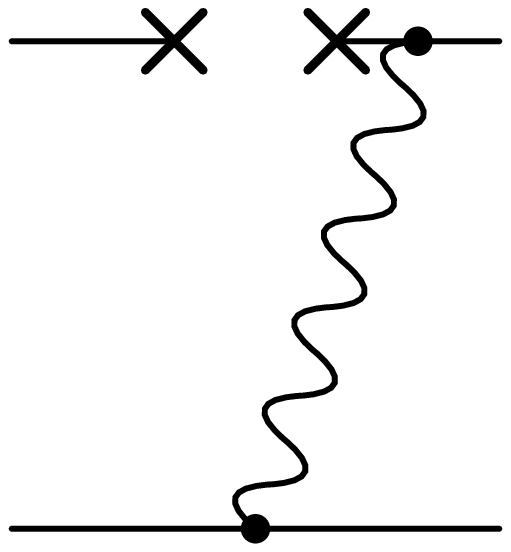, scale=0.5}\quad=:\quad
\epsfig{figure=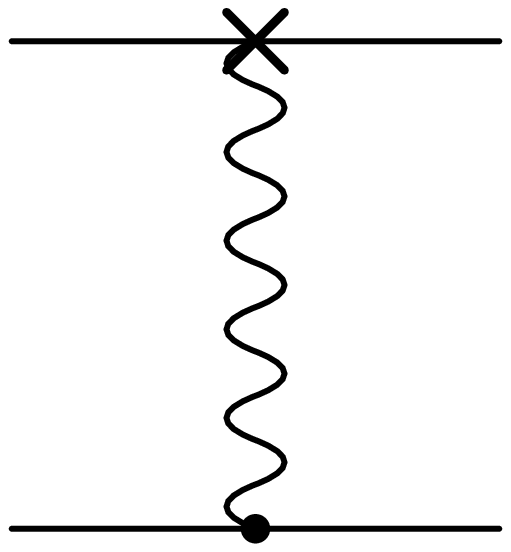, scale=0.5}\\[12pt]
(a)\kern300pt(b)\end{center}
\caption{\label{dicond}new modules for the correlator corrections including
  condensates: condensate correction (a) and condensate--propagator
  correction consisting of two diagrams~(b)}
\end{figure}

First we check on known results using configuration space techniques. We split
the result for the noncondensate contribution into two parts which reflect the
two ways of how the Dirac indices have been contracted. One has
\begin{eqnarray}\label{Pio}
\Pi^{\rm o}(x)&=&-4N_c!\left(F_0(x^2)\right)^3(1+t^2)
  \left\{1+\frac{\alpha_s}\pi\left(\left(\mu_x^2x^2\right)^\eps
  \left(\frac1\eps+\frac73\right)-\frac1\eps\right)\right\}x^2\slx\nonumber\\
  &=&-4N_c!\left(F_0(x^2)\right)^3(1+t^2)\left\{1+\frac{\alpha_s}\pi
  \left(\frac73+\ln\left(\mu_x^2x^2\right)\right)\right\}x^2\slx,\\
{\rm and} \nonumber\\\label{Pix}
\Pi^{\rm x}(x)&=&-N_c!\left(F_0(x^2)\right)^3(1+t)^2
  \left\{1+\frac{\alpha_s}\pi\left(\left(\mu_x^2x^2\right)^\eps
  \left(\frac1\eps+\frac73\right)-\left(\frac1\eps+\frac76\right)\right)
  \right\}x^2\slx\nonumber\\
  &=&-N_c!\left(F_0(x^2)\right)^3(1+t)^2\left\{1+\frac{\alpha_s}\pi
  \left(\frac76+\ln\left(\mu_x^2x^2\right)\right)\right\}x^2\slx
\end{eqnarray}
for the direct and crossover part of the Wick contraction, respectively, as
shown symbolically in Fig.~\ref{crossover}. The singular parts
$\propto 1/\eps$ in each of the contributions cancel against counter terms in
the course of renormalization performed in Eqs.~(\ref{Pio}) and~(\ref{Pix}).
Note the different dependence on the mixing parameter $t$ in the two parts of
$\Pi(x)=\Pi^{\rm o}(x)+\Pi^{\rm x}(x)$. Note also that our techniques allow
for the calculation of all condensate corrections but require new modules as
indicated in Fig.~\ref{dicond}. These modules are relevant for the calculation
of the coefficient functions of the scalar operator $\{\bar\psi\psi\}$ and the
vector operator $\{\bar\psi\gamma^\mu\psi\}$.

For the scalar condensate we obtain the results
\begin{eqnarray}
\Pi_S^{\rm o}(x)&=&4N_c!\left(F_0(x^2)\right)^2(1-t)(1+t)
  \left\{1+\frac{\alpha_s}\pi\left(\mu_x^2x^2\right)^\eps\right\}x^2
  \langle\bar\psi\psi\rangle/12 ,\nonumber\\
\Pi_S^{\rm x}(x)&=&N_c!\left(F_0(x^2)\right)^2(1-t)
  \left\{1+3t+\frac{\alpha_s}{2\pi}\left(\mu_x^2x^2\right)^\eps(1+7t)\right\}
  x^2\langle\bar\psi\psi\rangle/12 \nonumber.
\end{eqnarray}
where we have assumed $\langle\bar uu\rangle=\langle\bar dd\rangle
=\langle\bar\psi\psi\rangle$. Adding up the two contributions gives
\begin{equation}
\Pi_S(x)=N_c!\left(F_0(x^2)\right)^2(1-t)
  \left\{5+7t+\frac{3\alpha_s}{2\pi}\left(\mu_x^2x^2\right)^\eps(3+5t)\right\}
  x^2\langle\bar\psi\psi\rangle/12.
\end{equation}
These results are finite and need not be renormalized. After a Fourier
transformation the results are in agreement with the results for $\Pi_m(q^2)$
in Eq.~(\ref{Pimq2}) obtained by direct integration in momentum space. The
corresponding spectral density that appears in the integrand of the dispersion
representation reads
\begin{equation}
\rho_S(s)=\frac{N_c!s}{2(4\pi)^2}(1-t)\left\{5+7t+\frac{3\alpha_s}{2\pi}(3+5t)
  \right\}\langle\bar\psi\psi\rangle/12 
\end{equation}
with $s=q^2>0$. The result is proportional to $(1-t)$ and thus vanishes for
$t=1$. The vanishing of the spectral density at $t=1$ is a general property of
the correlator function related to the chiral structure of the current.
Indeed, in the massless limit where the chiral symmetry is exact the
contribution of the scalar quark condensate vanishes in all orders of
perturbation theory.

\begin{figure}\begin{center}
\epsfig{figure=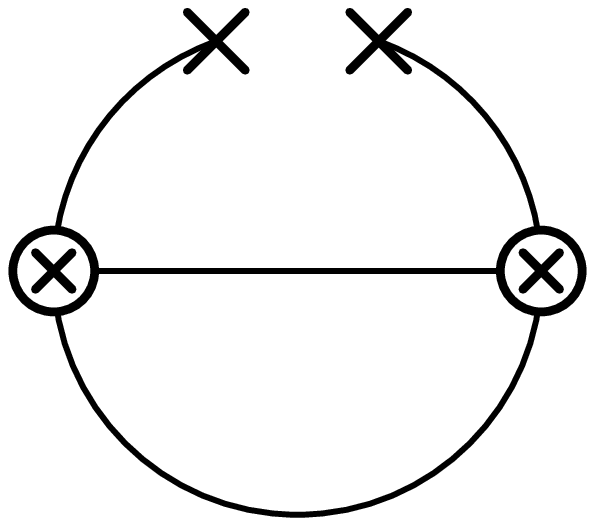, scale=0.4}\qquad
\epsfig{figure=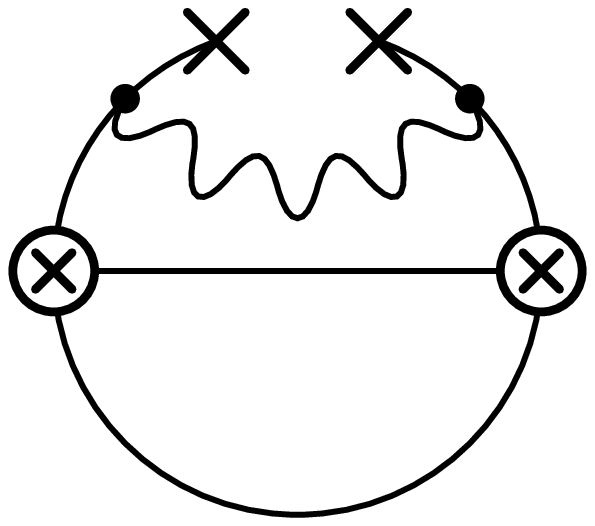, scale=0.4}\qquad
\epsfig{figure=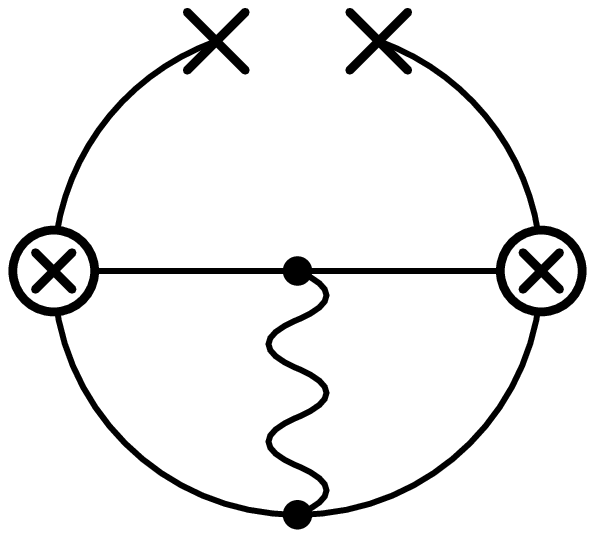, scale=0.4}\qquad
\epsfig{figure=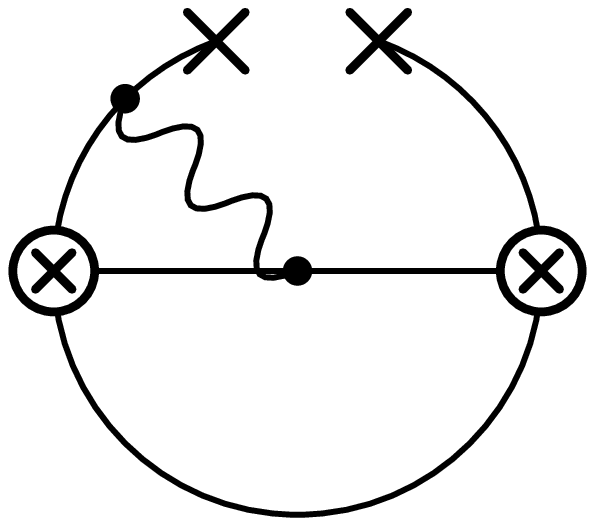, scale=0.4}\qquad
\epsfig{figure=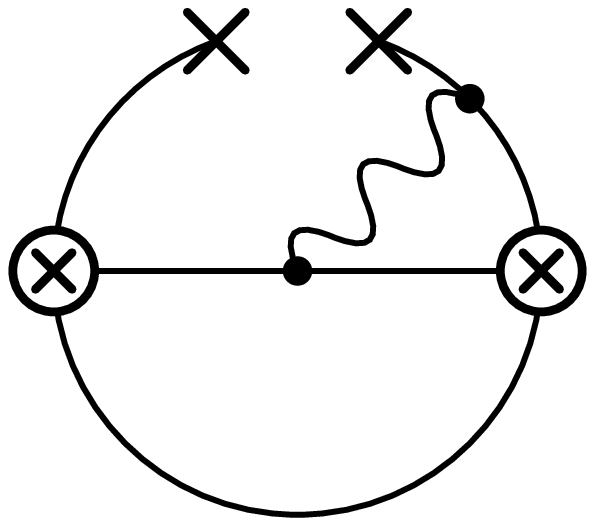, scale=0.4}\\[12pt]
(a)\kern76pt(b)\kern76pt(c)\kern76pt(d)\kern76pt(e)\end{center}
\caption{\label{conddia}LO contribution (a) and NLO contributions (b-e) to
  the correlation function including a (scalar or vector) condensate}
\end{figure}


\section{Correlator including the vector condensate}


We now present our new results for the vector condensate. As mentioned before
the vector condensate violates Lorentz invariance as a manifestation of the
presence of matter. It can be calculated by using the same set of diagrams as
before (see Fig.~\ref{conddia}). For the part proportional to the vector quark
operator in the OPE we write 
\begin{equation}
\Pi_V(x)=\frac{\{\bar\psi\gamma^\mu\psi\}}{12}
  \left(\slx x_\mu A_V(x^2)+x^2\gamma_\mu B_V(x^2)\right)
\end{equation}
where $A_V(x^2)=A_V^{\rm o}(x^2)+A_V^{\rm x}(x^2)$,
$B_V(x^2)=B_V^{\rm o}(x^2)+B_V^{\rm x}(x^2)$ and find the renormalized
coefficient functions
\begin{eqnarray}
A_V^{\rm o}(x^2)&=&-8N_c!\left(F_0(x^2)\right)^2(1+t^2)
  \left\{1+\frac{\alpha_s}\pi\left(\ln\left(\mu_x^2x^2\right)
  +\frac53\right)\right\},\nonumber\\
B_V^{\rm o}(x^2)&=&-4N_c!\left(F_0(x^2)\right)^2(1+t^2)
  \left\{1+\frac{\alpha_s}\pi\left(\ln\left(\mu_x^2x^2\right)
  +1\right)\right\},\nonumber\\
A_V^{\rm x}(x^2)&=&-2N_c!\left(F_0(x^2)\right)^2(1+t)^2
  \left\{1+\frac{\alpha_s}\pi\left(\ln\left(\mu_x^2x^2\right)
  +\frac12\right)\right\},\nonumber\\
B_V^{\rm x}(x^2)&=&-N_c!\left(F_0(x^2)\right)^2(1+t)^2
  \left\{1+\frac{\alpha_s}\pi\left(\ln\left(\mu_x^2x^2\right)
  -\frac16\right)\right\}.\qquad
\end{eqnarray}
Note that the curly bracket notation $\{\bar\psi\gamma^\mu\psi\}$ refers to an
operator before averaging. We emphasize again that the vacuum expectation
value of the vector operator vanishes, i.e.\
$\langle\bar\psi\gamma^\mu\psi\rangle=0$, while in matter one has
$\langle M|\bar\psi\gamma^\mu\psi|M\rangle\ne 0$.

In momentum space the correlator function is expanded as
\begin{equation}
\Pi_V(q)=\frac{\{\bar\psi\gamma^\mu\psi\}}{12}
  \left(\slq q_\mu A_V(q^2)+q^2\gamma_\mu B_V(q^2)\right).
\end{equation}
For the spectral density one obtains 
$\rho_{A_V}(s)=\rho_{A_V^{\rm o}}(s)+\rho_{A_V^{\rm x}}(s)$ and
$\rho_{B_V}(s)=\rho_{B_V^{\rm o}}(s)+\rho_{B_V^{\rm x}}(s)$ where
\begin{eqnarray}
\rho_{A_V^{\rm o}}(s)&=&\frac{4N_c!}{3(4\pi)^2}(1+t^2)\left\{1+
  \frac{\alpha_s}\pi\left(\frac72+\ell\right)\right\},\nonumber\\
\rho_{B_V^{\rm o}}(s)&=&-\frac{8N_c!}{3(4\pi)^2}(1+t^2)\left\{1+
  \frac{\alpha_s}\pi\left(\frac{15}4+\ell\right)\right\},
  \nonumber\\
\rho_{A_V^{\rm x}}(s)&=&\frac{N_c!}{3(4\pi)^2}(1+t)^2\left\{1+
  \frac{\alpha_s}\pi\left(\frac73+\ell\right)\right\},\nonumber\\
\rho_{B_V^{\rm x}}(s)&=&-\frac{2N_c!}{3(4\pi)^2}(1+t)^2\left\{1+
  \frac{\alpha_s}\pi\left(\frac{31}{12}+\ell\right)\right\}
\end{eqnarray}
and $\ell=\ln(\mu^2/s)$. As an example we take Ioffe's current
Eqs.~(\ref{ioffe1}) and~(\ref{ioffe2}) which is obtained from our general
current (\ref{general}) by setting $t=-1$ and multiplying by a factor of $2$.
Including the noncondensate contribution and the scalar and vector condensate
contributions the spectral density is now given by
\begin{eqnarray}
\rho_\eta(s)&=&\frac4{(4\pi)^4}\slq s^2
  \left\{1+\frac{\alpha_s}\pi\left(\frac{71}{12}+\ell\right)\right\}
  -\frac4{(4\pi)^2}\{\bar\psi\psi\}s\left\{1+\frac{3\alpha_s}{2\pi}\right\}
  \nonumber\\&&
  -\frac1{3\pi^2}\{\bar\psi\gamma^\mu\psi\}\left[\slq q_\mu\left\{1
  +\frac{\alpha_s}\pi\left(\frac72+\ell\right)\right\}
  +2s\gamma_\mu\left\{1+\frac{\alpha_s}{\pi}\left(\frac{15}4+\ell\right)
  \right\}\right]
\end{eqnarray}
Our results for the vector condensate confirm the LO results given
in~\cite{LO-prl,LO-druk}. The NLO corrections to the vector condensate are
new. One can see that they are numerically large in the $\MSbar$
renormalization scheme at the standard value $\mu=\sqrt s$ for the
renormalization scale. The numerical values of the condensates
$\langle M|\bar qq|M\rangle$ and $\langle M|\bar q\gamma_\mu q|M\rangle$ are
nonperturbative parameters of QCD that are built into the sum rule analysis.
Following~\cite{LO-prl,LO-druk} we take
$\langle M|\bar q\gamma_\mu q|M\rangle=u_\mu\frac32\rho_N$ where $u_{\mu}$ is
the four-velocity of relativistic nuclear matter and $\rho_N$ is its density.
For the contribution of the vector condensate we obtain
\begin{equation}
\rho_\eta^V(s)= -\frac1{2\pi^2}\rho_N\left[\slq(qu)\left\{1+\frac{\alpha_s}\pi
  \left(\frac72+\ell\right)\right\}
  +2s\slu\left\{1+\frac{\alpha_s}\pi\left(\frac{15}4+\ell\right)\right\}
  \right].
\end{equation}
Canonically QCD sum rules are analyzed at a low scale of the order of $1\GeV$.
The running of the coupling $\alpha_s(M_Z)=0.1176\pm 0.002$~\cite{PDG} to this
low scale $\mu=1\GeV$ results in $\alpha_s(1\GeV)/\pi=0.15\pm 0.1$. With this
value of the coupling constant the NLO correction amounts up to $60\%$ of the
leading order result. 

The inclusion of terms proportional to the light quark masses $m_{u,d}$ do not
substantially change the quantitative results as the masses of light quarks
are small~\cite{Gasser}. Even for exotic strange matter the results are still
valid since the $s$-quark mass is still reasonably small. Two recent
$O(\alpha_s^4)$ QCD sum rule determinations give
$m_s(2\GeV)=105\pm 6\pm 7\MeV$~\cite{CK} and
$m_s(2\GeV)=92\pm 9\MeV$~\cite{Jamin}. These numbers are rather close though
somewhat smaller than previous results based on $\tau$ decay
data~\cite{Korner:2000wd}. Even if there is not much hope to detect strange
matter on earth, strange matter can appear as an intermediate state in the
high energy collisions of heavy ions. In view of such possible applications
the inclusion of strange quark mass corrections is rather topical. We mention
that the contribution of four-quark operators are also
important~\cite{druk4quark}. Their contribution can be accounted for in the
factorization approximation. The result reads
\begin{eqnarray}
\Pi_{4q}(q^2)&=&-\frac{\langle\bar\psi\psi\rangle^2}{24Q^2}
  \Bigg(5\left\{1+\frac{\alpha_s}\pi\left(\frac{61}{15}L_Q-\frac{511}{90}
  \right)\right\}\\&&
  +2t\left(1+\frac{\alpha_s}\pi\left(\frac53L_Q-\frac{224}9\right)\right\}
  -7t^2\left\{1+\frac{\alpha_s}\pi\left(\frac{47}{21}L_Q+\frac{325}{126}
  \right)\right\}\Bigg)\nonumber
\end{eqnarray}
where $L_Q=\ln\pfrac{Q^2}{\mu^2}$. The accuracy of the factorization
approximation for four-quark operators has been checked in~\cite{Chetyrkin:yr}
where the configuration space technique was heavily used (see
also~\cite{Schafer:2000rv,Narison:2001ix}). 

\section{Conclusions}
To summarize, we found an important correction to the correlator of baryon
currents in media which is needed in the analysis of the properties of
relativistic nuclear matter and bound nucleons within the QCD sum rule
approach. The correction is given by the NLO contribution of QCD perturbation
theory expansion to the coefficient function of the vector condensate in the
OPE of the baryon currents and is not small. It amounts to $60\%$ of the
leading order term at a low energy scale relevant to the analysis of nuclear
matter and therefore should be taken into account in phenomenological
applications. 

\subsection*{Acknowledgements}
The work is supported by the Deutsche Forschungsgemeinschaft (DFG) under
contract No.~436~EST~17/1/06, by the RFFI grant No.~06-02-16353, by the
Estonian target financed project No.~0182647s04 and by the Estonian Science
Foundation under grant No.~6216. A.A.P. acknowledges the hospitality of the
Particle Theory Group at Siegen University where a part of this work was done
during his stay as a DFG Mercator Guest Professor
(DFG contract No.~SI~349/10-1).

\end{document}